\documentclass[aps,twocolumn,showpacs]{revtex4-1}
\usepackage{amsmath}
\usepackage{epsfig}
\usepackage{float}

\begin{document}

\title{Tetraquarks with open charm favor}

\newcommand*{\NJNU}{Department of Physics, Nanjing Normal University, Nanjing, Jiangsu 210097, China}\affiliation{\NJNU}

\author{Yaoyao Xue}\email[E-mail: ]{181002022@stu.njnu.edu.cn}
\author{Xin Jin}\email[E-mail: ]{181002005@stu.njnu.edu.cn}
\author{Hongxia Huang}
\email[E-mail: ]{hxhuang@njnu.edu.cn (Corresponding author)}
\author{Jialun Ping}
\email[E-mail: ]{jlping@njnu.edu.cn (Corresponding author)}
\affiliation{Department of Physics, Nanjing Normal University, Nanjing, Jiangsu 210097, China}

\begin{abstract}
Inspired by the recent report of the exotic states $X_{0}(2900)$ and $X_{1}(2900)$ with four different quark flavors in the $D^{-}K^{+}$ invariant mass distributions of the decay process $B^{+}\rightarrow D^{+}D^{-}K^{+}$ by the LHCb collaboration, we systematically investigate the tetraquarks composed of $ud\bar{s}\bar{c}$ with meson-meson and diquark-antidiquark structures in the quark delocalization color screening model. We find that the $X_{0}(2900)$ can be interpreted as the molecular state $\bar{D^{*}}K^{*}$ with $IJ^{P}=00^{+}$. Moreover, two bound states are obtained by the channel coupling calculation, with energies $2341.2$ MeV for $IJ^{P}=00^{+}$ and $2489.7$ MeV for $IJ^{P}=01^{+}$, respectively. We also extend our study to the $uc\bar{d}\bar{s}$ systems and find that there is no any bound state, so the $D_{s0}(2317)$ cannot be identified as the $DK$ molecular state in present calculation. Besides, several resonance states with the diquark-antidiquark configuration are possible in both $ud\bar{s}\bar{c}$ and $uc\bar{d}\bar{s}$ systems. All these open charm bound states and resonances are worth searching in the future experiments.
\end{abstract}


\maketitle

\setcounter{totalnumber}{5}

\section{\label{sec:introduction}Introduction}
Quest for exotic states beyond conventional hadron configurations is a long-standing challenge in hadron physics. So far, many tetraquark and pentaquark candidates are proposed, and most of them are composed of hidden charm or bottom quarks. Very recently, the LHCb collaboration reported the discovery of two new exotic structures $X_{0}(2900)$ and $X_{1}(2900)$ in the $D^{-}K^{+}$ invariant mass distributions of the decay process $B^{+}\rightarrow D^{+}D^{-}K^{+}$~\cite{LHCb1}. Since they are observed in the $D^{-}K^{+}$ channel, the lowest quark content of these two states should be $ud\bar{s}\bar{c}$, which imply that both $X_{0}(2900)$ and $X_{1}(2900)$ are possible to be open charm tetraquarks. Their spin-parity quantum numbers are $J^{P}=0^{+}$ and $1^{-}$, respectively, and masses and widths are:
\begin{eqnarray}
M_{X_{0}(2900)}&=&2.866\pm0.007~GeV, \nonumber \\
\Gamma_{X_{0}(2900)}&=& 57.2\pm 12.9 ~MeV, \nonumber \\
M_{X_{1}(2900)}&=&2.904\pm0.005~GeV, \nonumber \\
\Gamma_{X_{0}(2900)}&=& 110.3\pm 11.5 ~MeV, \nonumber
\end{eqnarray}

Motivated by the LHCb observation, a lot of theoretical works are proposed to explain these two exotic states~\cite{Karliner,MWHu,XGHe,XHLiu,JRZhang,QFLv,MZLiu,HXChen,JHe,ZGWang,YHuang}. In a very recent work of Ref.~\cite{Karliner}, the $X_{0}(2900)$ was interpreted as a $cs\bar{u}\bar{d}$ isosinglet compact tetraquark with the mass of $2863\pm12$ MeV. And the analogous $bs\bar{u}\bar{d}$ tetraquark was predicted at $6213\pm12$ MeV. These two open charm tetraquarks are also observed in the framework of QCD sum rules~\cite{JRZhang, HXChen,ZGWang}, the extended relativized quark model~\cite{QFLv}, one-boson exchange model~\cite{MZLiu,JHe}, and so on. The non-resonance explanation, triangle singularity mechanism, is also 
proposed~\cite{XHLiu}. Once the discovery of the  $X_{0}(2900)$ and $X_{1}(2900)$ is confirmed, a new exotic state with four different quark flavors 
will be attained and it will help us understanding the low-energy behavior of the QCD and the nature of the strong interactions.

In fact, the $X(5568)$ has been proposed as an exotic state with open flavors $us\bar{d}\bar{b}$ or $ds\bar{u}\bar{b}$ by D0 collaboration~\cite{D0}. Unfortunately, this state was not confirmed by other collaborations: the LHCb collaboration~\cite{LHCb2}, the CMS collaboration~\cite{CMS}, the CDF collaboration~\cite{CDF} and the ATLAS collaboration~\cite{ATLAS}. At the same time, another tetraquark with four different quark flavors $ud\bar{s}\bar{b}$ (or its charge-conjugated one) was proposed, which could be definitely observed via the weak decay mode $J/\psi K^{-}K^{-}\pi^{+}$~\cite{YuFS}. We have investigated tetraquarks composed of $us\bar{d}\bar{b}$ and $ud\bar{s}\bar{b}$ in the framework of the quark delocalization color screening model (QDCSM)~\cite{Huang1}, and found that the $X(5568)$ cannot be explained as a molecular state or a diquark-antidiquark resonance of $us\bar{d}\bar{b}$. Nevertheless, two tetraquarks composed of $ud\bar{s}\bar{b}$ were obtained with the diquark-antidiquark structure. So, the tetraquarks composed of $ud\bar{s}\bar{b}$ is more possible to form bound states than the one composed of $us\bar{d}\bar{b}$.

Because of the heavy favor symmetry, it is nature to extend the study to the tetraquarks composed of $ud\bar{s}\bar{c}$ and $uc\bar{d}\bar{s}$. The aims of this work are: (1) We study two structures of the open charm tetraquarks and to see if the newly reported $X_{0}(2900)$ and $X_{1}(2900)$ can be explained as the open charm tetraquarks in the constituent quark model, and explore the structure of these two states. (2) We do a systemically search of the open charm tetraquark systems to check if there is any other open charm tetraquarks. For example, the $D^{*}_{s0}(2317)$, first observed by the BaBar collaboration~\cite{Aubert}, appears as a very narrow resonance below the $DK$ threshold and decays to $D_{s}^{+}\pi^{0}$. One of the common interpretation is that it may be a $DK$ molecule state. Very recently, the lattice QCD observed the $DK$ and $D\bar{K}$ scattering process and found a near-threshold $IJ^{P}=00^{+}$ bound state $DK$, corresponding to the $D^{*}_{s0}(2317)$~\cite{lattic}. So it is also interesting to see whether there is any bound state below the threshold of $DK$, which may be used to explain the $D^{*}_{s0}(2317)$ in the quark approach.

The structure of the paper is as follows. A brief introduction of the quark model and wave functions are given in Sect. II. Sect. III is devoted to the numerical results and discussions. The summary is shown in the last section.

\section{MODEL AND WAVE FUNCTIONS}

\subsection{The model (QDCSM)}

The QDCSM has been widely described in the literatures~\cite{QDCSM0,QDCSM1}, and we refer the reader to those works for the details. 
Here, we just present the salient features of the model. The Hamiltonian of the QDCSM includes three parts: the rest masses of quarks, 
the kinetic energy and the interaction potentials. The potentials are composed of the color confinement (CON), the one-gluon exchange (OGE), 
and the one-Goldstone boson exchange (OBE). The detailed form for the tetraquark systems is shown below:
\begin{eqnarray}
H & = & \sum_{i=1}^4\left(m_i+\frac{p_i^2}{2m_i}\right)-T_{CM} +\sum_{j>i=1}^4 V_{ij} \\
V_{ij} & =& V^{\rm{CON}}_{ij}+V^{\rm{OGE}}_{ij}+V^{\rm{OBE}}_{ij} , \\
V_{ij}^{\rm{CON}} & = & -a_{c}\boldsymbol{\mathbf{\lambda}}^c_{i}\cdot
\boldsymbol{\mathbf{ \lambda}}^c_{j}~ (f(r_ij)+ a^{0}_{ij}), \\
f(r_{ij}) & = & \left \{ \begin{array}{ll}
r_{ij}^2, & \mbox{if \textit{i},\textit{j} in the same cluster} \\
\frac{1-e^{-\mu_{ij}\mathbf{r}_{ij}^2}}{\mu_{ij}}, & \mbox{otherwise}
\end{array} \right.\label{QDCSM-vc} \\
V^{\rm{OGE}}_{ij} & = & \frac{1}{4}\alpha_s \boldsymbol{\lambda}^{c}_i \cdot
\boldsymbol{\lambda}^{c}_j
\left[\frac{1}{r_{ij}}-\frac{\pi}{2}\delta(\boldsymbol{r}_{ij})\left(\frac{1}{m^2_i}+\frac{1}{m^2_j} \right. \right. \nonumber \\
& & ~~~~~~~~~~~~\left. \left. +\frac{4\boldsymbol{\sigma}_i\cdot\boldsymbol{\sigma}_j}{3m_im_j}\right)-\frac{3}{4m_im_jr^3_{ij}}
S_{ij}\right] \label{sala-vG} \\
V^{\rm{OBE}}_{ij} & = & V_{\pi}( \boldsymbol{r}_{ij})\sum_{a=1}^3\lambda
_{i}^{a}\cdot \lambda
_{j}^{a}+V_{K}(\boldsymbol{r}_{ij})\sum_{a=4}^7\lambda
_{i}^{a}\cdot \lambda _{j}^{a} \nonumber \\
&+ & V_{\eta}(\boldsymbol{r}_{ij})\left[\left(\lambda _{i}^{8}\cdot
\lambda _{j}^{8}\right)\cos\theta_P-(\lambda _{i}^{0}\cdot
\lambda_{j}^{0}) \sin\theta_P\right] \label{sala-Vchi1} \\
V_{\chi}(\boldsymbol{r}_{ij}) & = & {\frac{g_{ch}^{2}}{{4\pi
}}}{\frac{m_{\chi}^{2}}{{\
12m_{i}m_{j}}}}{\frac{\Lambda _{\chi}^{2}}{{\Lambda _{\chi}^{2}-m_{\chi}^{2}}}}%
m_{\chi} \Big\{ \Big[ Y(m_{\chi}\,r_{ij})  \nonumber \\
& & \left. -{\frac{\Lambda_{\chi}^{3}}{m_{\chi}^{3}}}Y(\Lambda _{\chi}\,r_{ij})\right] 
  \boldsymbol{\sigma}_{i}\cdot \boldsymbol{\sigma}_{j}
  +\Big[ H(m_{\chi} r_{ij})  \nonumber \\
& & \left. \left. -\frac{\Lambda_{\chi}^3}{m_{\chi}^3}
H(\Lambda_{\chi} r_{ij})\right] S_{ij} \right\}, ~~\chi=\pi, K, \eta, \\
S_{ij}&=&\left\{ 3\frac{(\boldsymbol{\sigma}_i
\cdot\boldsymbol{r}_{ij}) (\boldsymbol{\sigma}_j\cdot
\boldsymbol{r}_{ij})}{r_{ij}^2}-\boldsymbol{\sigma}_i \cdot
\boldsymbol{\sigma}_j\right\},\\
H(x)&=&(1+3/x+3/x^{2})Y(x),~~~
 Y(x) =e^{-x}/x. \label{sala-vchi2}
\end{eqnarray}
Where $S_{ij}$ is quark tensor operator; $Y(x)$ and $H(x)$ are
standard Yukawa functions; $T_c$ is the kinetic
energy of the center of mass; $\alpha_{s}$ is the quark-gluon coupling constant; $g_{ch}$ is the coupling constant
for chiral field, which is determined from the $NN\pi$ coupling constant through
\begin{equation}
\frac{g_{ch}^{2}}{4\pi }=\left( \frac{3}{5}\right) ^{2}{\frac{g_{\pi NN}^{2}%
}{{4\pi }}}{\frac{m_{u,d}^{2}}{m_{N}^{2}}}\label{gch}.
\end{equation}
The other symbols in the above expressions have their usual meanings. All model parameters are determined by fitting the meson
spectrum and shown in Table~\ref{parameters}. The calculated masses of the mesons in comparison with experimental values are shown 
in Table~\ref{mass}.
\begin{table}[h]
\caption{\label{parameters}Model parameters:
$m_{\pi}=0.7$ fm$^{-1}$, $m_{ K}=2.51$ fm$^{-1}$,
$m_{\eta}=2.77$ fm$^{-1}$, $\Lambda_{\pi}=4.2$ fm$^{-1}$, $\Lambda_{K}=\Lambda_{\eta}=5.2$ fm$^{-1}$,
$g_{ch}^2/(4\pi)=0.54$, $\theta_p = -15^{0}$. }
\begin{tabular}{ccccccccc} \hline\hline
$b$ & ~~~$m_{u}$~~~~ & ~~~$m_{d}$~~~ & ~~~$m_{s}$~~~ & ~~~$m_{c}$~~~&~~~$m_{b}$&~~~~   \\
(fm) & (MeV) & (MeV) & (MeV) & (MeV) &(MeV)    \\ \hline
0.518  & 313 & 313 &  470  & 1270 &4500\\ \hline
$ a_c$ &  $a^{0}_{uu}$ &  $a^{0}_{us}$ & $a^{0}_{uc}$ & $a^{0}_{sc}$ & $a^{0}_{ub}$ & $a^{0}_{sb}$ \\
(MeV\,fm$^{-2}$) & (fm$^{2}$) & (fm$^{2}$) & (fm$^{2}$) & (fm$^{2}$) & (fm$^{2}$)  & (fm$^{2}$) \\ \hline
  58.03 & -0.733 & -0.309 & 1.278 & 1.358  & 1.701 & 1.808 \\ \hline
 $\alpha_{s_{uu}}$ &  $\alpha_{s_{us}}$ & $\alpha_{s_{uc}}$ & $\alpha_{s_{sc}}$ & $\alpha_{s_{ub}}$ & $\alpha_{s_{sb}}$  \\ \hline
  1.50 & 1.46 & 1.450 & 1.44& 1.41 & 1.40 \\
 \hline\hline
\end{tabular}
\end{table}

\begin{table}[ht]
\caption{The masses (in MeV) of the mesons obtained from QDCSM. Experimental values are taken
from the Particle Data Group (PDG)~\cite{PDG}.}
\begin{tabular}{lccccccccc}
\hline \hline
 ~~~~~Meson~~~~~& ~~~~~$M_{the}$~~~~~ & ~~~~~$M_{exp}$~~~~~ \\ \hline
 ~~~~~~~~$\pi$     & 140 & 140 \\
 ~~~~~~~~$\rho$    & 772 & 770 \\
 ~~~~~~~~$D$       & 1865  &1869   \\
 ~~~~~~~~$D^{*}$   & 2008  & 2008  \\
 ~~~~~~~~$D_{s}$   & 1968   & 1968  \\
 ~~~~~~~~$D^{*}_{s}$   & 2062  & 2112 \\
 ~~~~~~~~$K$       & 495 & 495 \\
 ~~~~~~~~$K^{*}$   & 892 & 892 \\
 ~~~~~~~~$B$       & 5280 & 5280 \\
 ~~~~~~~~$B^{*}$   & 5319 & 5325 \\
 ~~~~~~~~$B_{s}$   & 5367 & 5367 \\
 ~~~~~~~~$B^{*}_{s}$  & 5393 & 5415 \\   \hline\hline

\end{tabular}
\label{mass}
\end{table}

The quark delocalization in QDCSM is realized by specifying the
single particle orbital wave function of QDCSM as a linear
combination of left and right Gaussians, which are the single particle
orbital wave functions used in the ordinary quark cluster model,
\begin{eqnarray}
\psi_{\alpha}(\mathbf{s}_i ,\epsilon) & = & \left(
\phi_{\alpha}(\mathbf{s}_i)
+ \epsilon \phi_{\alpha}(-\mathbf{s}_i)\right) /N(\epsilon), \nonumber \\
\psi_{\beta}(-\mathbf{s}_i ,\epsilon) & = &
\left(\phi_{\beta}(-\mathbf{s}_i)
+ \epsilon \phi_{\beta}(\mathbf{s}_i)\right) /N(\epsilon), \nonumber \\
N(\epsilon) & = & \sqrt{1+\epsilon^2+2\epsilon e^{-s_i^2/4b^2}}. \label{1q} \\
\phi_{\alpha}(\mathbf{s}_i) & = & \left( \frac{1}{\pi b^2}
\right)^{3/4}
   e^{-\frac{1}{2b^2} (\mathbf{r}_{\alpha} - \mathbf{s}_i/2)^2} \nonumber \\
\phi_{\beta}(-\mathbf{s}_i) & = & \left( \frac{1}{\pi b^2}
\right)^{3/4}
   e^{-\frac{1}{2b^2} (\mathbf{r}_{\beta} + \mathbf{s}_i/2)^2}. \nonumber
\end{eqnarray}
Here $\mathbf{s}_i$, $i=1,2,...,n$ are the generating coordinates, which are introduced to expand the relative motion
wave function~\cite{QDCSM1}. The delocalization parameter $\epsilon(\mathbf{s}_i)$ is not an adjustable one but determined
variationally by the dynamics of a multi-quark system itself. In this way, the multi-quark system can choose its favorable 
configuration in a larger Hilbert space.

\subsection{Wave functions}
In this work, we study the tetraquark systems in two structures: the meson-meson structure and the diquark-antidiquark structure. The resonating group method (RGM)~\cite{RGM}, a well-established method for studying a bound-state or a scattering problem, is used to calculate the energy of all these states. The wave function of the four-quark system is of the form
\begin{equation}
\Psi = {\cal A } \left[[\psi^{L}\psi^{\sigma}]_{JM}\psi^{f}\psi^{c}\right],
\end{equation}
where $\psi^{L}$, $\psi^{\sigma}$, $\psi^{f}$, and $\psi^{c}$ are the orbital, spin, flavor and color wave functions respectively, which are shown below. The symbol ${\cal A }$ is the anti-symmetrization operator. For the meson-meson structure of $u\bar{s}-d\bar{c}$, ${\cal A }$ is defined as
\begin{equation}
{\cal A } = 1-P_{13},
\end{equation}
for the $u\bar{d}-c\bar{s}$
\begin{equation}
{\cal A } = 1-P_{24},
\end{equation}
and for the diquark-antidiquark structure $ud-\bar{s}\bar{c}$
\begin{equation}
{\cal A }=1-P_{12}.
\end{equation}
The orbital wave function is the same in two configurations, and the spin wave functions is the same too. But the flavor and color wave functions are constructed differently depending on different structures.

\subsubsection{The orbital wave function}

The orbital wave function is in the form of
\begin{equation}
\psi^{L} = {\psi}_{1}(\boldsymbol{R}_{1}){\psi}_{2}(\boldsymbol{R}_{2})\chi_{L}(\boldsymbol{R}).
\end{equation}
where $\boldsymbol{R}_{1}$ and $\boldsymbol{R}_{2}$ are the internal coordinates for the cluster 1 and cluster 2. $\boldsymbol{R} = \boldsymbol{R}_{1}-\boldsymbol{R}_{2}$ is the relative coordinate between the two clusters 1 and 2. The ${\psi}_{1}$ and ${\psi}_{2}$ are 
the internal cluster orbital wave functions of the clusters 1 and 2, which are fixed in the calculation, and $\chi_{L}(\boldsymbol{R})$ is 
the relative motion wave function between two clusters, which is expanded by gaussian bases
\begin{eqnarray}
& & \chi_{L}(\boldsymbol{R}) = \frac{1}{\sqrt{4\pi}}(\frac{3}{2\pi b^2}) \sum_{i=1}^{n} C_{i}  \nonumber \\
&& ~~~~\times  \int \exp\left[-\frac{3}{4b^2}(\boldsymbol{R}-\boldsymbol{s}_{i})^{2}\right] Y_{LM}(\hat{\boldsymbol{s}_{i}})d\hat{\boldsymbol{s}_{i}}. ~~~~~
\end{eqnarray}
where $\boldsymbol{s}_{i}$ is called the generate coordinate, $n$ is the number of the gaussian bases, which is determined by the stability of the results. By doing this, the integro-differential equation of RGM can be reduced to an algebraic equation, generalized eigen-equation. Then we can obtain the energy of the system by solving this generalized eigen-equation. The details of solving the RGM equation can be found in Ref.~\cite{RGM}.

\subsubsection{ The flavor wave function}
For the meson-meson configuration, as the first step, we give the wave functions of the meson cluster, which are shown below.
\begin{eqnarray}
\chi^{I1}_{00} &=& \frac{1}{\sqrt{2}}(u\bar{u}+d\bar{d}),~~~~\chi^{I2}_{10} = \frac{1}{\sqrt{2}}(d\bar{d}-u\bar{u}), \nonumber \\
\chi^{I3}_{\frac{1}{2}\frac{1}{2}} &=& u\bar{c},~~~~\chi^{I4}_{00} = c\bar{s}, ~~~~\chi^{I5}_{\frac{1}{2}\frac{1}{2}} = u\bar{s},\nonumber\\
\chi^{I6}_{\frac{1}{2}\frac{1}{2}} &=& c\bar{d},~~~~\chi^{I7}_{\frac{1}{2}-\frac{1}{2}}=d\bar{s}.~~~~\chi^{I8}_{\frac{1}{2}-\frac{1}{2}} = d\bar{c} , \nonumber \\
\chi^{I9}_{\frac{1}{2}-\frac{1}{2}} &=& - c\bar{u}
\end{eqnarray}
where the superscript of the $\chi$ is the index of the flavor wave function for a meson, and the subscript stands for the isospin  $I$ and the third component $I_{z}$.
The flavor wave functions with the meson-meson structure are:
\begin{eqnarray}
\psi^{f_{1}}_{00} &=& \chi^{I4}_{00}\chi^{I1}_{00},~~~~\psi^{f_{2}}_{11} = \chi^{I4}_{10}\chi^{I2}_{10},  \nonumber \\
\psi^{f_{3}}_{00} &=& \sqrt{\frac{1}{2}}\left[\chi^{I6}_{\frac{1}{2}\frac{1}{2}}\chi^{I7}_{\frac{1}{2}-\frac{1}{2}}-\chi^{I9}_{\frac{1}{2}-\frac{1}{2}}\chi^{I5}_{\frac{1}{2}\frac{1}{2}}\right],  \nonumber \\
\psi^{f_{4}}_{11} &=& \sqrt{\frac{1}{2}}\left[\chi^{I6}_{\frac{1}{2}\frac{1}{2}}\chi^{I7}_{\frac{1}{2}-\frac{1}{2}}+\chi^{I9}_{\frac{1}{2}-\frac{1}{2}}\chi^{I5}_{\frac{1}{2}\frac{1}{2}}\right],
\nonumber \\
\psi^{f_{5}}_{00} &=& \sqrt{\frac{1}{2}}\left[\chi^{I3}_{\frac{1}{2}\frac{1}{2}}\chi^{I7}_{\frac{1}{2}-\frac{1}{2}}-\chi^{I8}_{\frac{1}{2}\frac{1}{2}}\chi^{I5}_{\frac{1}{2}\frac{1}{2}}\right],  \nonumber \\
\psi^{f_{6}}_{11} &=& \sqrt{\frac{1}{2}}\left[\chi^{I3}_{\frac{1}{2}\frac{1}{2}}\chi^{I7}_{\frac{1}{2}-\frac{1}{2}}+\chi^{I8}_{\frac{1}{2}\frac{1}{2}}\chi^{I5}_{\frac{1}{2}\frac{1}{2}}\right].
\end{eqnarray}

For the diquark-antidiquark configuration, we first show the functions of the diquark and antidiquark, respectively.
\begin{eqnarray}
\chi^{I1}_{{10}} &=& \frac{1}{\sqrt{2}}(ud+du),~~~~\chi^{I2}_{{00}} = \frac{1}{\sqrt{2}}(ud-du),  \nonumber \\
\chi^{I3}_{{\frac{1}{2}\frac{1}{2}}} &=& cu,~~~~\chi^{I4}_{{\frac{1}{2}-\frac{1}{2}}} = cd,  \nonumber \\
\chi^{I5}_{{\frac{1}{2}-\frac{1}{2}}} &=& -\bar{s}\bar{u},~~~~\chi^{I6}_{{\frac{1}{2}\frac{1}{2}}} = \bar{s}\bar{d},\nonumber \\
\chi^{I7}_{{00}} &=& \bar{c}\bar{s}.
\end{eqnarray}
Then, the flavor wave functions for the diquark-antidiquark structure can be obtained by coupling the wave functions of two clusters.
\begin{eqnarray}
\psi^{f_{1}}_{00} &=& \sqrt{\frac{1}{2}}\left[\chi^{I3}_{{\frac{1}{2}\frac{1}{2}}}\chi^{I5}_{{\frac{1}{2}-\frac{1}{2}}}-\chi^{I4}_{{\frac{1}{2}-\frac{1}{2}}}\chi^{I6}_{{\frac{1}{2}\frac{1}{2}}}\right],  \nonumber \\
\psi^{f_{2}}_{10} &=& \sqrt{\frac{1}{2}}\left[\chi^{I3}_{{\frac{1}{2}\frac{1}{2}}}\chi^{I5}_{{\frac{1}{2}-\frac{1}{2}}}+\chi^{I4}_{{\frac{1}{2}-\frac{1}{2}}}\chi^{I6}_{{\frac{1}{2}\frac{1}{2}}}\right],  \nonumber \\
\psi^{f_{3}}_{11} &=& \chi^{I2}_{{00}}\chi^{I7}_{{00}},~~~~\psi^{f_{4}}_{00} = \chi^{I1}_{{10}}\chi^{I7}_{{00}}.
\end{eqnarray}

\subsubsection{ The spin wave function}
The spin wave function of a meson cluster is:
\begin{eqnarray}
\chi^{\sigma 1}_{{11}} &=& \alpha\alpha,~~~~\chi^{\sigma 2}_{{10}} = \sqrt{\frac{1}{2}}(\alpha\beta+\beta\alpha), \nonumber \\
~~~~\chi^{\sigma 3}_{{1-1}} &=& \beta\beta,~~~~\chi^{\sigma4}_{{00}} = \sqrt{\frac{1}{2}}(\alpha\beta-\beta\alpha).
\end{eqnarray}
Then the spin wave functions of the four-quark system are:
\begin{eqnarray}
\psi^{\sigma_{1}}_{00} &=& \chi^{\sigma 4}_{{00}}\chi^{\sigma 4}_{{00}},\nonumber \\
\psi^{\sigma_{2}}_{00} &=& \sqrt{\frac{1}{3}}\left[\chi^{\sigma 1}_{{11}}\chi^{\sigma 3}_{{1-1}}-\chi^{\sigma 2}_{{10}}\chi^{\sigma 2}_{{10}}+\chi^{\sigma 3}_{{1-1}}\chi^{\sigma 1}_{{11}}\right],  \nonumber \\
\psi^{\sigma_{3}}_{11} &=& \chi^{\sigma 4}_{{00}}\chi^{\sigma 1}_{{11}},~~~~\psi^{\sigma_{4}}_{11} = \chi^{\sigma 1}_{{11}}\chi^{\sigma 4}_{{00}},  \nonumber \\
\psi^{\sigma_{5}}_{11} &=& \sqrt{\frac{1}{2}}\left[\chi^{\sigma 1}_{{11}}\chi^{\sigma 2}_{{10}}-\chi^{\sigma 2}_{{10}}\chi^{\sigma 1}_{{11}}\right], \nonumber \\
\psi^{\sigma_{6}}_{22} &=& \chi^{\sigma 1}_{{11}}\chi^{\sigma 1}_{{11}}.
\end{eqnarray}

\subsubsection{The color wave function}
The color wave function of a meson cluster is:
\begin{eqnarray}
\chi^{1}_{[111]} &=& \sqrt{\frac{1}{3}}(r\bar{r}+g\bar{g}+b\bar{b}).
\end{eqnarray}
and the four-quark system wave function with the meson-meson structure is
\begin{eqnarray}
\psi^{c_{1}} &=& \chi^{1}_{[111]}\chi^{1}_{[111]}.
\end{eqnarray}

For the diquark-antidiquark structure, the color wave functions of the diquark clusters are:
\begin{eqnarray}
\chi^{1}_{[2]} &=& rr,~~~\chi^{2}_{[2]} = \frac{1}{\sqrt{2}}(rg+gr),~~~\chi^{3}_{[2]} = gg,  \nonumber \\
\chi^{4}_{[2]} &=& \frac{1}{\sqrt{2}}(rb+br),~~~\chi^{5}_{[2]} = \frac{1}{\sqrt{2}}(gb+bg),~~~\chi^{6}_{[2]} = bb,  \nonumber \\
\chi^{7}_{[11]}&=& \frac{1}{\sqrt{2}}(rg-gr),~~~\chi^{8}_{[11]}= \frac{1}{\sqrt{2}}(rb-br), \nonumber \\
\chi^{9}_{[11]}&=& \frac{1}{\sqrt{2}}(gb-bg).
\end{eqnarray}
and the color wave functions of the antidiquark clusters are:
\begin{eqnarray}
\chi^{1}_{[22]} &=& \bar{r}\bar{r},~~~\chi^{2}_{[22]} = -\frac{1}{\sqrt{2}}(\bar{r}\bar{g}+\bar{g}\bar{r}),~~~~\chi^{3}_{[22]} = \bar{g}\bar{g},  \nonumber \\
\chi^{4}_{[22]} &=& \frac{1}{\sqrt{2}}(\bar{r}\bar{b}+\bar{b}\bar{r}),~\chi^{5}_{[22]} = -\frac{1}{\sqrt{2}}(\bar{g}\bar{b}+\bar{b}\bar{g}),~\chi^{6}_{[22]} = \bar{b}\bar{b},  \nonumber \\
\chi^{7}_{[211]}&=& \frac{1}{\sqrt{2}}(\bar{r}\bar{g}-\bar{g}\bar{r}),~~~~\chi^{8}_{[211]}= -\frac{1}{\sqrt{2}}(\bar{r}\bar{b}-\bar{b}\bar{r}), \nonumber \\
\chi^{9}_{[211]}&=& \frac{1}{\sqrt{2}}(\bar{g}\bar{b}-\bar{b}\bar{g}).
\end{eqnarray}
Then, the wave functions for the four-quark system with the diquark-antidiquark structure can be obtained by coupling the wave functions of the diquark and antidiquark clusters, which are:
\begin{eqnarray}
\psi^{c_{1}} &=& \sqrt{\frac{1}{6}}[\chi^{1}_{[2]}\chi^{1}_{[22]}-\chi^{2}_{[2]}\chi^{2}_{[22]}+\chi^{3}_{[2]}\chi^{3}_{[22]}  \nonumber \\
&& +\chi^{4}_{[2]}\chi^{4}_{[22]}-\chi^{5}_{[2]}\chi^{5}_{[22]}+\chi^{6}_{[2]}\chi^{6}_{[22]}],  \nonumber \\
\psi^{c_{2}} &=& \sqrt{\frac{1}{3}}\left[\chi^{7}_{[11]}\chi^{7}_{[211]}-\chi^{8}_{[11]}\chi^{8}_{[211]}+\chi^{9}_{[11]}\chi^{9}_{[211]}\right].
\end{eqnarray}
Finally, multiplying the wave functions $\psi^{L}$, $\psi^{\sigma}$, $\psi^{f}$, and $\psi^{c}$ according to the definite quantum number of the system, we can acquire the total wave functions.

\section{The results and discussions}
In present work, we investigate tetraquarks with two kinds of quark components: $ud\bar{s}\bar{c}$ and $uc\bar{d}\bar{s}$. Two structures, meson-meson and diquark-antidiquark, are considered. The quantum numbers of the tetraquarks we study here are $I=0,~1$, $J=0,~1,~2$ and the parity is $P=+$. All the orbital 
angular momenta are set to zero because we are interested in the ground states in this work. To check whether or not there is any bound state in such 
tetraquark system, we do a dynamic bound-state calculation. Both the single-channel and channel-coupling calculations are carried out in this work. 
All the general features of the calculated results are as follows.

\subsection{Tetraquarks $ud\bar{s}\bar{c}$}
For tetraquarks composed of $ud\bar{s}\bar{c}$, the possible quantum numbers can be $IJ=00$, $01$, $02$, $10$, $11$ and $12$. The energies of the meson-meson structure are listed in Table~\ref{bound1}, where the second column gives the index of the wave functions of every channel, and the wave functions are from Eqs. (17), (21), and (23). The third column is the corresponding channel. $E_{th}$ denotes the theoretical threshold of every channel; $E_{sc}$ and $E_{cc}$ represent the energies of the single-channel and channel-coupling calculation respectively.

\begin{table}
\caption{The energies (in MeV) of the meson-meson structure for tetraquarks $ud\bar{s}\bar{c}$.}
\begin{tabular}{lcccccccccc}
\hline \hline
  & ~$[\psi^{f_{i}}\psi^{\sigma_{j}}\psi^{c_{k}}]$~ & ~Channel~ & ~$E_{th}$~~ & ~~$E_{sc}$~~ & ~~$E_{cc}$~~  \\ \hline
 ~~$IJ=00$~~ & $[\psi^{f_{5}}\psi^{\sigma_{1}}\psi^{c_{1}}]$ & $\bar{D}K$ & $2360.0$ & $2367.0$ & $2341.2$ \\
             & $[\psi^{f_{5}}\psi^{\sigma_{2}}\psi^{c_{1}}]$ & $\bar{D^{*}}K^{*}$ & $2900.5$ & $2820.7$ &  \\ \hline
 ~~$IJ=01$~~ & $[\psi^{f_{5}}\psi^{\sigma_{4}}\psi^{c_{1}}]$ & $\bar{D^{*}}K$ & $2503.0$ & $2509.6$ & $2489.7$\\
             & $[\psi^{f_{5}}\psi^{\sigma_{3}}\psi^{c_{1}}]$ & $\bar{D}K^{*}$ & $2757.5$ & $2761.3$ &  \\
             & $[\psi^{f_{5}}\psi^{\sigma_{5}}\psi^{c_{1}}]$ & $\bar{D^{*}}K^{*}$ & $2900.5$ & $2904.2$ &  \\\hline
 ~~$IJ=02$~~ & $[\psi^{f_{5}}\psi^{\sigma_{5}}\psi^{c_{1}}]$ & $\bar{D^{*}}K^{*}$ & $2900.5$ & $2908.4$ & $$ \\ \hline
 ~~$IJ=10$~~ & $[\psi^{f_{6}}\psi^{\sigma_{1}}\psi^{c_{1}}]$ & $\bar{D}K$ & $2360.0$ & $2369.5$ & $2369.3$ \\
             & $[\psi^{f_{6}}\psi^{\sigma_{2}}\psi^{c_{1}}]$ & $\bar{D^{*}}K^{*}$ & $2900.5$ & $2907.6$ &  \\ \hline
 ~~$IJ=11$~~ & $[\psi^{f_{6}}\psi^{\sigma_{4}}\psi^{c_{1}}]$ & $\bar{D^{*}}K$ & $2503.0$ & $2511.0$ & $2511.0$ \\
             & $[\psi^{f_{6}}\psi^{\sigma_{3}}\psi^{c_{1}}]$ & $\bar{D}K^{*}$ & $2757.5$ & $2766.5$ &  \\
             & $[\psi^{f_{6}}\psi^{\sigma_{5}}\psi^{c_{1}}]$ & $\bar{D^{*}}K^{*}$ & $2900.5$ & $2908.0$ &  \\ \hline
~~$IJ=12$~~ & $[\psi^{f_{6}}\psi^{\sigma_{5}}\psi^{c_{1}}]$ & $\bar{D^{*}}K^{*}$ & $2900.5$ & $2905.3$ & $$ \\
 \hline\hline
\end{tabular}
\label{bound1}
\end{table}

From the Table~\ref{bound1}, we can see that the energies of every single channel are above the corresponding theoretical threshold, except the $\bar{D^{*}}K^{*}$ state with $IJ=00$. The energy of this state is $2820.7$ MeV, about $80$ MeV lower than the threshold of $\bar{D^{*}}K^{*}$. Since the energy is close to the newly reported $X_{0}(2900)$ and the spin-parity quantum numbers are $J^{P}=0^{+}$, which is also consistent with $X_{0}(2900)$, it is reasonable to identify the $X_{0}(2900)$ as a molecular state $\bar{D^{*}}K^{*}$ with $IJ^{P}=00^{+}$ in our quark model calculation.

We also investigate the effect of the multi-channel coupling. It is obvious in Table~\ref{bound1} that two bound states are obtained after channel coupling calculation. One  is the tetraquark state with $IJ=00$, the energy of which is $2341.2$ MeV, almost $20$ MeV lower than the threshold of $\bar{D}K$. Although the energy of this state is close to the mass of $D^{*}_{s0}(2317)$, it cannot be used to explain the $D^{*}_{s0}(2317)$. Because the quark component here is $ud\bar{s}\bar{c}$, it cannot decay to the $D_{s}^{+}\pi^{0}$ channel. Another one is the tetraquark state with $IJ=01$, the energy of which is $2489.7$ MeV, $13.3$ MeV lower than the threshold of $\bar{D^{*}}K$. In the same way, it cannot be used to identify $D_{s1}(2460)$, though the energy is close to the $D_{s1}(2460)$. The states with other quantum numbers are unbound after the channel coupling, which indicates that the effect of the channel coupling for these systems is very small and cannot help much. Therefore, the channel coupling plays an important role in forming bound states for both the $IJ=00$ and $IJ=01$ tetraquark systems, while it can be neglected for the systems with other quantum numbers.

With regard to the tetraquarks in diquark-antidiquark structure, the energy of each single channel is higher than the theoretical threshold of the corresponding channel, which are shown in Table~\ref{bound2}. After the channel coupling calculation, the energy of the $IJ=00$ system was pushed down to $2206.7$ MeV, $153$ MeV lower than the theoretical threshold, which indicates that the $IJ=00$ state of the diquark-antidiquark structure is possible to be a bound state. For the systems with other quantum numbers, although the effect of channel-coupling is much stronger than that of the meson-meson structure, the energy is still above the theoretical threshold of the corresponding channel. So there is no any bound states for the systems with $IJ=01$, $02$, $10$, $11$ or $IJ=12$ in the diquark-antidiquark structure. However, it is possible for them to be resonance states, because the colorful subclusters diquark ($ud$) and antidiquark ($\bar{s}\bar{c}$) cannot fall apart due to the color confinement. To check the possibility, we carry out an adiabatic calculation of the effective potentials for the $ud\bar{s}\bar{c}$ system with diquark-antidiquark structure.

\begin{table}
\caption{The energies (in MeV) of the diquark-antidiquark structure for tetraquarks $ud\bar{s}\bar{c}$.}
\begin{tabular}{lcccccccccc}
\hline \hline
  & ~$[\psi^{f_{i}}\psi^{\sigma_{j}}\psi^{c_{k}}]$~ &~~~~$E_{th}$~~~~ &  ~~~~$E_{sc}$~~~~ & ~~~~$E_{cc}$~~~~  \\ \hline
~~$IJ=00$~~ & $[\psi^{f_{4}}\psi^{\sigma_{1}}\psi^{c_{2}}]$  & $2360.0$&2512.4 & $2206.7$ \\
             & $[\psi^{f_{4}}\psi^{\sigma_{2}}\psi^{c_{1}}]$  & &2575.8 &  \\ \hline
~~$IJ=01$~~ & $[\psi^{f_{4}}\psi^{\sigma_{3}}\psi^{c_{2}}]$  & $2503.0$ &$2572.2$& $2534.7$ \\
             & $[\psi^{f_{4}}\psi^{\sigma_{4}}\psi^{c_{1}}]$  &  &$3023.4$&  \\
             & $[\psi^{f_{4}}\psi^{\sigma_{5}}\psi^{c_{1}}]$  & &$2825.4$&  \\ \hline
~~$IJ=02$~~ & $[\psi^{f_{4}}\psi^{\sigma_{1}}\psi^{c_{1}}]$  & $2900.5$ & $3130.7$ \\ \hline
~~$IJ=10$~~ & $[\psi^{f_{3}}\psi^{\sigma_{2}}\psi^{c_{2}}]$  & $2360.0$ &$2851.5$& $2519.7$ \\
             & $[\psi^{f_{3}}\psi^{\sigma_{1}}\psi^{c_{1}}]$  &  &3147.1&  \\ \hline
~~$IJ=11$~~ & $[\psi^{f_{3}}\psi^{\sigma_{3}}\psi^{c_{2}}]$  & $2503.0$ & $2928.6$ &$2807.5$\\
             & $[\psi^{f_{3}}\psi^{\sigma_{4}}\psi^{c_{2}}]$  &  &$2912.7$&  \\
             & $[\psi^{f_{3}}\psi^{\sigma_{5}}\psi^{c_{1}}]$  & &$3130.6$&  \\ \hline
~~$IJ=12$~~ & $[\psi^{f_{3}}\psi^{\sigma_{1}}\psi^{c_{2}}]$  & $2900.5$ & $3013.7$ \\
 \hline\hline
\end{tabular}
\label{bound2}
\end{table}

The effective potential is obtained by the formula $V_{E}(S)=E(S)-E_{th}$, where $E_{th}$ is the threshold of the corresponding lowest channel, and $E(S)$ is the energy at each $S$, which is the distance between two subclusters. Here $E(S)$ is obtained by:
\begin{eqnarray}
E(S)=\frac{\langle \Psi (S)|H|\Psi (S)\rangle}{\langle \Psi (S)|\Psi (S)\rangle}.  \nonumber
\end{eqnarray}
where $\langle \Psi (S)|H|\Psi (S)\rangle$ and $\langle \Psi (S)|\Psi (S)\rangle$ are the Hamiltonian matrix and the overlap of the state.
The effective potentials as a function of the distance between the diquark and antidiquark for the $ud\bar{s}\bar{c}$ system are shown in Fig.~1, where $sc1$, $sc2$ and $sc3$ stand for the potential of the first, the second and the third single channel respectively shown in Table~\ref{bound2}.

\begin{figure}[ht]
\begin{center}
\epsfxsize=3.5in \epsfbox{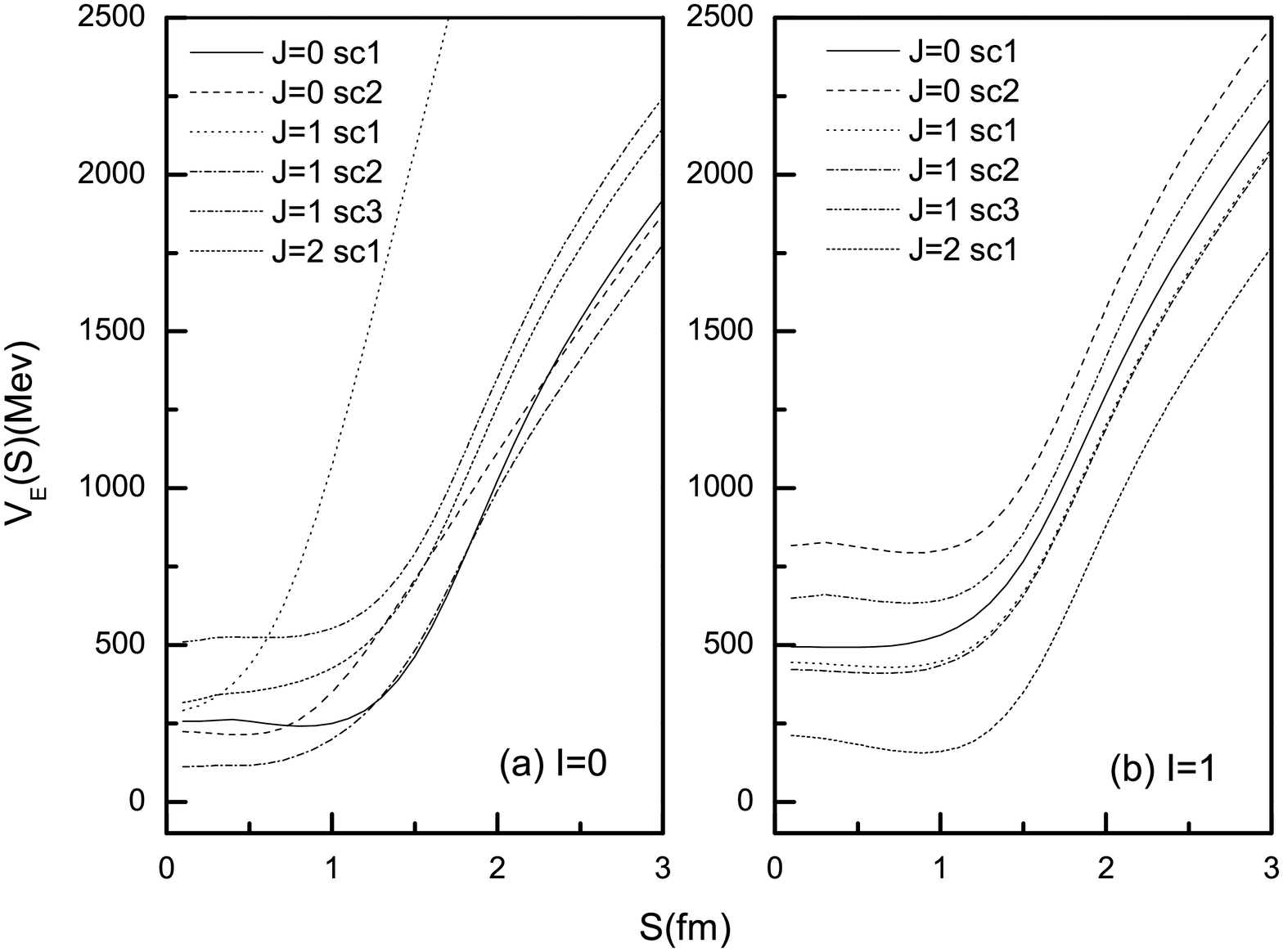} \vspace{-0.4in}

\caption{The effective potentials as a function of the distance between the diquark ($ud$) and antidiquark ($\bar{s}\bar{c}$) for the $ud\bar{s}\bar{c}$ system.}
\end{center}
\end{figure}

From the Fig.~1(a) we can see that the effective potential of each channel with $IJ=01$ is increasing when the two subclusters fall apart, which means
that the diquark and antidiquark tend to clump together without hinderance. This behavior indicates that the odds for the states being
diquark-antidiquark structure, meson-meson structure, or other structures are the same. Besides, from the Tables~\ref{bound1} and \ref{bound2}, we can see that the energy of the each channel with $IJ=01$ in the diquark-antidiquark structure is higher than the one in the meson-meson structure. So the state prefers to be two free mesons.
Therefore, none of these states is a observable resonance state in present calculation. It is different for the channels with other quantum numbers, where the energy of the state will rise when the two subclusters are too close, so there is a hinderance for the states of diquark-antidiquark structure changing to meson-meson structure even if the energy of the state is lower in meson-meson structure. Therefore, it is possible to form a wide resonance. The resonance energies are about $2500 \sim 3100$ MeV.
However, all these states will couple to the open channels. To confirm whether the states can survive as a resonance state after coupling to the open channels, further study of the scattering process of the open channels is needed in future work.
In addition, among all these resonance states, we notice that the energy of a $IJ=11$ resonance is $2912.7$ MeV, which is close to the newly reported $X_{1}(2900)$. However, the spin-parity quantum numbers are $J^{P}=1^{+}$, which is opposite to the experimental data $1^{-}$. Therefore, it may not be used to explain the $X_{1}(2900)$ state here. The tetraquark systems with $P-$wave should be considered to observe the exotic state $X_{1}(2900)$.

\subsection{Tetraquarks $uc\bar{d}\bar{s}$}
For tetraquarks composed of $uc\bar{d}\bar{s}$, the energies of the meson-meson structure and the diquark-antidiquark structure are listed in Table~\ref{bound3} and \ref{bound4}, respectively. For the meson-meson configuration, we can see from the Table~\ref{bound3} that the energies of every single channel approach to the corresponding theoretical threshold, which means that there is no any bound state for every single channel. The channel coupling effect is very small and cannot help much, and the energies are still higher than the theoretical thresholds, which indicates that no any bound state of the $uc\bar{d}\bar{s}$ system in the meson-meson structure is formed in our quark model calculation.
Particularly, the $DK$ state is unbound here, which shows that the $D^{*}_{s0}(2317)$ cannot be identified as the $DK$ molecular state in present calculation.

\begin{table}
\caption{The energies (in MeV) of the meson-meson structure for tetraquarks $uc\bar{d}\bar{s}$.}
\begin{tabular}{lcccccccccc}
\hline \hline
  & ~$[\psi^{f_{i}}\psi^{\sigma_{j}}\psi^{c_{k}}]$~ & ~Channel~ & ~$E_{th}$~~ & ~~$E_{sc}$~~ & ~~$E_{cc}$~~  \\ \hline
~~$IJ=00$~~ & $[\psi^{f_{1}}\psi^{\sigma_{1}}\psi^{c_{1}}]$ & $D_{s}\eta$ & $2252.3$ & $2259.9$ & $2256.6$ \\
             & $[\psi^{f_{1}}\psi^{\sigma_{2}}\psi^{c_{1}}]$ & $D_{s}\omega$ & $2787.0$ & $2791.1$ &  \\
             & $[\psi^{f_{3}}\psi^{\sigma_{1}}\psi^{c_{1}}]$ & $DK$ & $2360.0$ & $2368.6$ &  \\
             & $[\psi^{f_{3}}\psi^{\sigma_{2}}\psi^{c_{1}}]$ & $D^{*}K^{*}$ & $2900.5$ & $2907.5$ &  \\ \hline
~~$IJ=01$~~ & $[\psi^{f_{1}}\psi^{\sigma_{4}}\psi^{c_{1}}]$ & $D^{*}_{s}\eta$ & $2346.9$ & $2352.5$ & $2347.0$ \\
             & $[\psi^{f_{1}}\psi^{\sigma_{3}}\psi^{c_{1}}]$ & $D_{s}\omega$ & $2692.4$ & $2698.1$ &  \\
             & $[\psi^{f_{1}}\psi^{\sigma_{5}}\psi^{c_{1}}]$ & $D^{*}\omega$ & $2787.0$ & $2791.2$ &  \\
             & $[\psi^{f_{3}}\psi^{\sigma_{4}}\psi^{c_{1}}]$ & $D^{*}K$ & $2503.0$ & $2510.2$ &  \\
              & $[\psi^{f_{3}}\psi^{\sigma_{3}}\psi^{c_{1}}]$ & $DK^{*}$ & $2757.5$ & $2765.0$ &  \\
             & $[\psi^{f_{3}}\psi^{\sigma_{5}}\psi^{c_{1}}]$ & $D^{*}K^{*}$ & $2900.5$ & $2907.2$ &  \\ \hline
~~$IJ=02$~~ & $[\psi^{f_{1}}\psi^{\sigma_{5}}\psi^{c_{1}}]$ & $D^{*}_{s}\omega$ & $2787.0$ & $2791.4$ & $2790.9$ \\
             & $[\psi^{f_{3}}\psi^{\sigma_{5}}\psi^{c_{1}}]$ & $D^{*}K^{*}$ & $2900.5$ & $2905.5$ &  \\ \hline
~~$IJ=10$~~ & $[\psi^{f_{2}}\psi^{\sigma_{1}}\psi^{c_{1}}]$ & $D_{s}\pi$ & $2108.1$ & $2116.0$ & $2114.6$ \\
             & $[\psi^{f_{2}}\psi^{\sigma_{2}}\psi^{c_{1}}]$ & $D^{*}_{s}\rho$ & $2835.1$ & $2838.6$ &  \\
             & $[\psi^{f_{4}}\psi^{\sigma_{1}}\psi^{c_{1}}]$ & $DK$ & $2360.0$ & $2368.6$ &  \\
             & $[\psi^{f_{4}}\psi^{\sigma_{2}}\psi^{c_{1}}]$ & $D^{*}K^{*}$ & $2900.5$ & $2906.0$ &  \\\hline
~~$IJ=11$~~ & $[\psi^{f_{2}}\psi^{\sigma_{4}}\psi^{c_{1}}]$ & $D^{*}_{s}\pi$ & $2202.7$ & $2209.6$ & $2208.2$ \\
             & $[\psi^{f_{2}}\psi^{\sigma_{3}}\psi^{c_{1}}]$ & $D_{s}\rho$ & $2740.5$ & $2746.0$ &  \\
             & $[\psi^{f_{2}}\psi^{\sigma_{5}}\psi^{c_{1}}]$ & $D^{*}_{s}\rho$ & $2835.1$ & $2838.8$ &  \\
             & $[\psi^{f_{4}}\psi^{\sigma_{4}}\psi^{c_{1}}]$ & $D^{*}K$ & $2503.0$ & $2510.2$ &  \\
             & $[\psi^{f_{4}}\psi^{\sigma_{3}}\psi^{c_{1}}]$ & $DK^{*}$ & $2757.5$ & $2765.0$ &  \\
             & $[\psi^{f_{4}}\psi^{\sigma_{5}}\psi^{c_{1}}]$ & $D^{*}K^{*}$ & $2900.5$ & $2906.4$ &  \\\hline

~~$IJ=12$~~ & $[\psi^{f_{2}}\psi^{\sigma_{5}}\psi^{c_{1}}]$ & $D^{*}_{s}\rho$ & $2835.1$ & $2839.0$ & $2836.5$ \\
             & $[\psi^{f_{4}}\psi^{\sigma_{5}}\psi^{c_{1}}]$ & $D^{*}K^{*}$ & $2900.5$ & $2906.8$ &  \\
 \hline\hline
\end{tabular}
\label{bound3}
\end{table}

For the diquark-antidiquark structure, it is obvious that the energy of every system is much higher than that of the meson-meson structure. Thus, there is no bound state with diquark-antidiquark structure. To check if there is any resonance state, we also perform an adiabatic calculation of the effective potentials for this $uc\bar{d}\bar{s}$ system with diquark-antidiquark structure, which are shown in Fig.~2. It is clear that the variation tendency of the potentials of the $uc\bar{d}\bar{s}$ system with $I=0$ or $I=1$ is similar to the one of the $ud\bar{s}\bar{c}$ system with $I=1$. The energy of the state increases a little when the two subclusters get too close, which causes a hinderance for the states changing structure to two mesons even if the energy of the diquark-antidiquark structure is higher than that of the  meson-meson structure. As a result, it is possible to form wide resonance states here, with resonance energies from about $2580$ MeV to $3100$ MeV.
All these resonances should be checked further by coupling to the open channels.

\begin{table}
\caption{The energies (in MeV) of the diquark-antidiquark structure for tetraquarks $uc\bar{d}\bar{s}$.}
\begin{tabular}{lcccccccccc}
\hline \hline
  & ~$[\psi^{f_{i}}\psi^{\sigma_{j}}\psi^{c_{k}}]$~~~~&~~~~$E_{th}$~~~~ & ~~~~$E_{sc}$~~~~ & ~~~~$E_{cc}$~~~~  \\ \hline
~~$IJ=00$~~ & $[\psi^{f_{1}}\psi^{\sigma_{2}}\psi^{c_{2}}]$ &2360.0 & $2939.9$ & $2712.9$ \\
             & $[\psi^{f_{1}}\psi^{\sigma_{1}}\psi^{c_{1}}]$ & & $2965.7$ &  \\ \hline
~~$IJ=01$~~ & $[\psi^{f_{1}}\psi^{\sigma_{3}}\psi^{c_{2}}]$ &2503.0 & $2938.3$ & $2735.3$ \\
             & $[\psi^{f_{1}}\psi^{\sigma_{5}}\psi^{c_{2}}]$ & & $2970.6$ &  \\
             & $[\psi^{f_{1}}\psi^{\sigma_{4}}\psi^{c_{1}}]$ & & $2938.1$ &  \\ \hline
~~$IJ=02$~~ & $[\psi^{f_{1}}\psi^{\sigma_{5}}\psi^{c_{2}}]$ &2900.5 & $3024.3$ & $$ \\ \hline
~~$IJ=10$~~ & $[\psi^{f_{2}}\psi^{\sigma_{1}}\psi^{c_{1}}]$ &2360.0 & $2850.0$ & $2584.3$ \\
             & $[\psi^{f_{2}}\psi^{\sigma_{2}}\psi^{c_{2}}]$ && $2965.7$ &  \\ \hline
~~$IJ=11$~~ & $[\psi^{f_{2}}\psi^{\sigma_{3}}\psi^{c_{2}}]$ &2503.0 & $2938.3$ & $2675.2$ \\
             & $[\psi^{f_{2}}\psi^{\sigma_{5}}\psi^{c_{2}}]$ & & $2930.2$ &  \\
             & $[\psi^{f_{2}}\psi^{\sigma_{4}}\psi^{c_{1}}]$ & & $2938.1$ &  \\  \hline
~~$IJ=12$~~ & $[\psi^{f_{2}}\psi^{\sigma_{5}}\psi^{c_{2}}]$ &2900.5& $3057.7$ & $$ \\
 \hline\hline
\end{tabular}
\label{bound4}
\end{table}

\begin{figure}[ht]
\begin{center}
\epsfxsize=3.5in \epsfbox{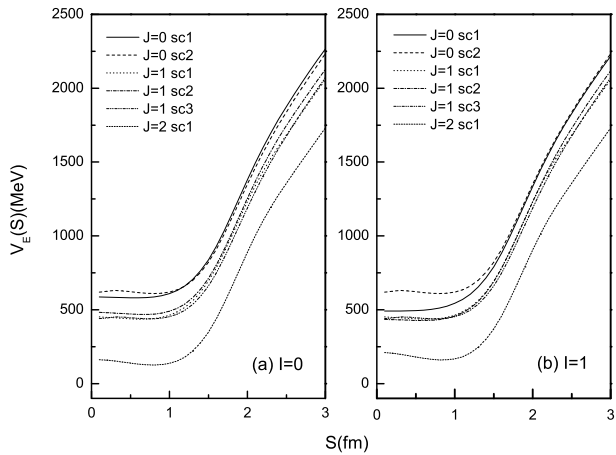} \vspace{-0.4in}

\caption{The effective potentials as a function of the distance between the diquark ($uc$) and antidiquark ($\bar{d}\bar{s}$) for the $uc\bar{d}\bar{s}$ system.}
\end{center}
\end{figure}

\section{Summary}
In this work, we systematically observe the $S-$wave tetraquarks composed of $ud\bar{s}\bar{c}$ and $uc\bar{d}\bar{s}$ in the framework of QDCSM.
Two structures, meson-meson and diquark-antidiquark, are considered. A dynamic bound-state calculation is carried out to look for bound states in such open charm tetraquark systems. In the calculation, both the single-channel and the channel-coupling are implemented. Besides, an adiabatic calculation of the effective potentials is justify the possibility of any resonance state.

For the $ud\bar{s}\bar{c}$ systems: (1) The dynamical calculation shows that there is a bound state $\bar{D^{*}}K^{*}$ with energy $2820.7$ MeV and quantum numbers $IJ^{P}=00^{+}$, which is possible to explain the newly reported $X_{0}(2900)$. However, the $\bar{D^{*}}K^{*}$ can decay to the $\bar{D}K$ channel. To confirm whether the states of $\bar{D^{*}}K^{*}$ can survive as a resonance state after coupling to the scattering state, further study of the scattering process of $\bar{D}K$ is needed. The energy of the $\bar{D^{*}}K^{*}$ state would be inflated by coupling to the open channel, much closer to the mass of the $X_{0}(2900)$.
Besides, two bound states $\bar{D}K$ and $\bar{D^{*}}K$ are obtained by the channel coupling calculation. Their energies are $2341.2$ MeV and $2489.7$ MeV, with quantum numbers $IJ^{P}=00^{+}$ and $IJ^{P}=01^{+}$, respectively. (2) The effective potentials of the diquark-antidiquark structure indicate several wide resonances are possible in present calculation, with the resonance energies about $2500 \sim 3100$ MeV. Among all these resonance states, we find the energy of a resonance with $IJ^{P}=11^{+}$ is $2912.7$ MeV, which is close to the newly reported $X_{1}(2900)$, but the parity is opposite to the reported one. So, it may not be used to explain the $X_{1}(2900)$ state here.

For the $uc\bar{d}\bar{s}$ systems: (1) The dynamical calculation shows that there is no any bound state in the meson-meson structure or the diquark-antidiquark structure.
So the $D_{s0}(2317)$ cannot be identified as the $DK$ molecular state in present calculation. (2) The effective potentials of the diquark-antidiquark structure also show the possibility of several wide resonances in the quark model calculation, with the resonance energies about $2580 \sim 3100$ MeV.

We study the open charm tetraquarks in two structures in this work. There are also other structures, e.g., $K$-type structure. Structure mixing may
lower the energies of the systems. However this is not a easy job. The over-completeness problem has to be solved. In addition, to confirm the existence of 
the resonances with open charm, the study of the scattering process of the corresponding open channels is needed. All these open charm bound states and 
resonances are worth searching in the future experiments.

\acknowledgments{This work is supported partly by the National Science Foundation
of China under Contract Nos. 11675080, 11775118 and 11535005}.


\begin{thebibliography}{99}
\bibitem{LHCb1} LHC Seminar, $B\rightarrow D\bar{D}h$ decays: A new (virtual) laboratory for exotic particle searches at LHCb, by Daniel Johnson, CERN, August 11, 2020, https://indico.cern.ch/event/900975/ .
\bibitem{Karliner} M. Karliner and J. L. Rosner, arXiv:2008.05993 [hep-ph].
\bibitem{MWHu} M. W. Hu, X. Y. Lao, P. Ling, and Q. Wang, arXiv:2008.06894 [hep-ph].
\bibitem{XGHe} X. G. He, W. Wang, and R. L. Zhu, arXiv:2008.07145 [hep-ph].
\bibitem{XHLiu} X. H. Liu, M. J. Yan, H. W. Ke, G. Li, and J. J. Xie, arXiv:2008.0719 [hep-ph].
\bibitem{JRZhang} J. R. Zhang, arXiv:2008.07295 [hep-ph].
\bibitem{QFLv} Q. F. Lu, D. Y. Chen, and Y. B. Dong, arXiv:2008.07340 [hep-ph].
\bibitem{MZLiu} M. Z. Liu, J. J. Xie, and L. S. Geng, arXiv:2008.07389 [hep-ph].
\bibitem{HXChen} H. X. Chen, W. Chen, R. R. Dong, and N. Su, arXiv:2008.07516 [hep-ph].
\bibitem{JHe} J. He and D. Y. Chen, arXiv:2008.07782 [hep-ph].
\bibitem{ZGWang} Z. G. Wang, arXiv:2008.07833 [hep-ph].
\bibitem{YHuang} Y. Huang, J. X. Lu, J. J. Xie, and L. S. Geng, arXiv:2008.07959 [hep-ph].
\bibitem{D0} V. M. Abazov {\em et al.} (D0 Collaboration), Phys. Rev. Lett. {\bf 117}, 022003 (2016).
\bibitem{LHCb2} R. Aaij {\em et al.} (LHCb Collaboration), Phys. Rev. Lett. {\bf 117}, 152003 (2016).
\bibitem{CMS} A. M. Sirunyan {\em et al.} (CMS Collaboration), Phys. Rev. Lett. {\bf 120}, 202005 (2018).
\bibitem{CDF} T. A. Aaltonen {\em et al.} (CDF Collaboration), Phys. Rev. Lett. {\bf 120}, 202006 (2018).
\bibitem{ATLAS} M. Aaboud {\em et al.} (ATLAS Collaboration), Phys. Rev. Lett. {\bf 120}, 202007 (2018).
\bibitem{YuFS} F. S. Yu, arXiv:1709.02571 [hep-ph].
\bibitem{Huang1} H. X. Huang and J. L. Ping, Eur. Phys. J. C {\bf 79}, 556 (2019).
\bibitem{Aubert}B. Aubert \textit{et al.} (BaBar Collaboration), Phys. Rev. Lett. {\bf 90}, 242001 (2003).
\bibitem{lattic} G. K. C. Cheung, C. E. Thomas, D. J. Wilson, G. Moir, M. Peardon, and S. M. Ryan, arXiv:2008.06432 [hep-lat].
\bibitem{QDCSM0} F. Wang, G. H. Wu, L. J. Teng and T. Goldman, Phys. Rev. Lett. {\bf 69}, 2901 (1992).
\bibitem{QDCSM1} J. L. Ping, F. Wang, and T. Goldman, Nucl. Phys. A {\bf 657}, 95 (1999).
\bibitem{PDG} Particle Data Group, C. Patrignani {\em et al.}, Chin. Phys. C {\bf 40}, 100001 (2016).
\bibitem{RGM} M. Kamimura, Prog. Theor. Phys. Suppl. {\bf 62}, 236 (1977).
\end{thebibliography}
\end{document}